\documentclass[preprint,showpacs]{revtex4}
\usepackage{graphics,graphicx,epsfig}

\begin{document}

\title{Thin film growth by random deposition of linear polymers on a square lattice}

\author{F. L. Forgerini}
\email{fabricio_forgerini@ufam.edu.br}
\affiliation{ISB, Universidade Federal do Amazonas, 69460-000 Coari-AM, Brazil}

\author{W. Figueiredo}
\email{wagner@fisica.ufsc.br}
\affiliation{Departamento de F\'{i}sica, Universidade Federal de Santa Catarina,
88040-900 Florian\'opolis-SC, Brazil }

\date{\today}
\begin{abstract}
We present some results of Monte Carlo simulations for the deposition of particles of different sizes on a two-dimensional substrate. The particles are linear, height one, and can be deposited randomly only in the two, $x$ and $y$ directions of the substrate, and occupy an integer number of cells of the lattice. We show there are three different regimes for the temporal evolution of the interface width. At the initial times we observe an uncorrelated growth, with an exponent $\beta_{1}$ characteristic of the random deposition model. At intermediate times, the interface width presents an unusual behavior, described by a growing exponent $\beta_{2}$, which depends on the size of the particles added to the substrate. If the linear size of the particle is two we have $\beta_{2}<\beta_{1}$, otherwise we have $\beta_{2}>\beta_{1}$, for all other particle sizes. After a long time the growth reaches the saturation regime where the interface width becomes constant and is described by the roughness exponent $\alpha$, which is nearly independent of the size of the particle. Similar results are found in the surface growth due to the electrophoretic deposition of polymer chains. Contrary to one-dimensional results the growth exponents are non-universal.
\end{abstract}

\pacs{68.43.Jk ; 68.35. Ct ; 81.15.Aa ; 82.35.Gh ; 02.50.-r}

\maketitle

\section{Introduction}

The formation of structures due to the deposition of particles is a topic of growing interest, and has challenged theoretical and experimental researchers in the field of material physics \cite{barabasi,meakin}. From the experimental point of view there is a real possibility of developing thin film devices with important technological applications and, at the same time, the theoretical physicists can apply the tools well known in the realm of statistical equilibrium physics to describe these new nonequilibrium phenomena.

It is well established that some characteristics of the growing surfaces present scale-invariant properties, so that quite different growing processes exhibit very similar scaling behavior, which we believe is universal. The temporal evolution of a surface formed by the deposition of particles is usually described in terms of some scale exponents. These scaling exponents define the most fundamental characteristics of the growth processes, so that we can put into the same universality class many different processes that have the same values of the scaling exponents.

In this work we are interested in the description of the morphology of a surface formed by adding particles to an initially flat two-dimensional substrate. Particles having one unit height and linear size $N$ land horizontally onto the surface and are not allowed to diffuse. In this way, we calculate the interface width, $w(L,t)$, a function that determines the roughness of the interface, where $L$ is the side of a two-dimensional square substrate and $t$ is time variable. In order to calculate the interface width, we determine the vertical height of a given point of the surface relative to the substrate, $h(\textbf{r},t)$, where \textbf{r} gives the position of the cells on the substrate. The roughness $w(L,t)$ is defined as the mean square fluctuation of the height, $w(L,t)=\langle[h(\textbf{r},t)-\overline{h}(t)]^{2}\rangle^{1/2}$, where $\overline{h}(t)$ is the average value of the surface height at a given instant of time $t$.

For a large number of growth models, Family and Vicsek proposed a scaling relation, relating the surface roughness with the linear size of the lattice and time \cite{FV}. This scaling relation is written as

\begin{equation}
w(L,t)\sim L^{\alpha}f(\frac{t}{L^{z}}),
\label{family_vicsek}
\end{equation}
where the scaling function $f(x)$ is a constant when $x$ is very large and $f(x)\sim x^{\beta}$ when $x<<1$. The exponent $\alpha$ is related to the saturation of the interface width at long times, $z$ is called the dynamic exponent and $\beta$ is the exponent that
measures the evolution of the interface width at the initial times of deposition. They are not independent, and are related by the equation

\begin{equation}
\beta=\frac{\alpha}{z}.
\label{alpha_beta_z}
\end{equation}

The kinetics of growing interfaces can also be studied by means of stochastic differential equations, which describe the evolution, in space and time, of the points at the growing surface. These equations capture the essence of the discrete models and serve to put them in
the proper universality class. The most important universality classes are defined by the Edwards-Wilkinson (EW) equation \cite{EW82}:

\begin{equation}
\frac{\partial h({\textbf{r}},t)}{\partial t}=\nu\nabla^{2}h+\eta({\textbf{r}},t) 
\label{EW}
\end{equation}
and by the Kardar-Parisi-Zhang (KPZ) equation \cite{KPZ86}:

\begin{equation}
\frac{\partial h({\textbf{r}},t)}{\partial t}=\nu{\nabla}^{2}h+\frac{\lambda}{2}(\nabla h)^{2}+\eta({\textbf{r}},t),
\label{KPZ}
\end{equation}
where $\eta({\textbf{r}},t)$ represents the random fluctuations in the deposition processes, with zero configurational average, and uncorrelated in space and time.

Generally, the linear equation (EW) represents random deposition models, where correlations between nearest neighbor sites are present, as for instance, models including surface relaxation. An important generalization of the EW equation is the KPZ equation that includes a non-linear term $\lambda(\nabla h)^{2}$ in the EW equation. It takes into account
lateral correlations and it is useful to describe growth processes occuring along the local normal to the surface. It represents a wide variety of processes of surface growth and non-equilibrium interfaces, such as those related to the formation of porous surfaces \cite{Vold}, corrosion processes of metallic surfaces \cite{reis_stafiej}, dissolution of a crystalline solid in a liquid medium \cite{Fernando}, etc.

There are many studies in the literature concerning the properties of the growth models where calculations have been done analytically, by solving the stochastic differential equations, employing mean-field calculations, or through the extensive use of Monte Carlo simulations
\cite{das_sarma,albano,drossel,horowitz_arvia,Reis,barato_oliveira}. The simplest non-trivial known discrete models are the random deposition model with surface relaxation and the ballistic deposition model, described by the equations EW and KPZ, respectively.

Discrete atomistic models that are commonly presented in the literature, are in general related to the deposition of identical particles, with the same size as the lattice parameter of the substrate. In some studies, different models of particle deposition are combined \cite{cerdeira_95,elnassar_cerdeira}, or even different species of particles are deposited at the same time, a feature observed in some systems \cite{karmakar,trojan,caglioti}. Interfaces generated by the deposition of particles larger than one cell of the underlying substrate have been considered, especially for models describing the growth surface due to polymer chains deposition \cite{bentrem,seung-woo}. Some other important questions related to the thin-film growth have received some attention recently, such as the lattice geometry and temperature effects \cite{gangshi}.

Recently, we have investigated the surface growth generated by the random deposition of particles of different sizes on a one-dimensional substrate \cite{forgerini_figueiredo}. We have shown that the roughness, growth and dynamic exponents of the model are independent of the size of the particles. The results of our simulations have shown that the roughness evolves in time following three different behaviors. The roughness at the initial times behaves as in the random deposition model. At intermediate times, the surface roughness grows slowly, and finally, at long times, it enters into the saturation regime. The scaling exponents of the model are the same as those predicted by the Villain-Lai-Das Sarma equation.

In the present study we investigate the properties of the model in two dimensions. Particles can be deposited independently in the $x$ and $y$ directions of the substrate. We also observe three different regimes for the temporal evolution of the interface width. However, contrary to the findings in one dimension, the growing exponent determined at intermediate times depends strongly on the size of the particle that is considered for deposition. For particles whose size is larger than two units of the lattice parameter of the substrate, this exponent is always larger than the one observed at the initial times of deposition. While in one dimension a new particle can be accomodated in one of the two ends of an already deposited particle, independently of its size, in two dimensions, the new deposition can be done in the $2(1+N)$ positions around a deposited particle, where $N$ is the linear size of the particle. This dependence on $N$ makes the growth in two dimensions to be non-universal.

This paper is organized as follows. In Sec. II we give a brief description of the model, along with the aggregation rules for the deposition of linear N-mers. In Sec. III we describe the details involved in the Monte Carlo simulations, and in Sec. IV we present the results of our simulations for the scaling exponents, showing the anomalous behavior of the growth exponent as a function of the size of the particles at intermediate times. Finally, in Sec. V, we present our conclusions.

\section{Model description}

We consider in this work a simple model that can mimic the growth of thin films due to the deposition of linear N-mers on a flat substrate. In this model the particles are dropped randomly over a square lattice containing $L^{2}$ unit cells. All particles to be deposited are one unit height and the columns where they land are increased by at least one unit, because voids are created during the deposition process. At the beginning of the deposition, we assume that we have a flat surface, and that particles land always horizontally on the surface. As we will see next, after some initial steps in the growth, the interface width behaves in an unusual way depending strongly on the size of the particle that is being deposited.

The particles deposited on the substrate are linear N-mers of size (1 X 1 X N), with $1\leq N\leq15$. They are added to the substrate in accordance with the rules of the random deposition model. A particle is aggregated only if the site for deposition coincides with the midpoint of the particle and there is enough space to accomodate it in the $x$ or $y$ directions. Otherwise, if the cell is already occupied or there is no enough space for the deposition, the particle is reflected away due to these geometric constraints. Particles are not allowed to diffuse or share the same cell in any deposition plane. As we will see next, during a unit of time, we try to deposit $L^{2}/N$ particles, which roughly means the deposition of one layer. Therefore, depending on the size of the particle, many trials of deposition are not successful in the unit of time. Other possible mechanisms of deposition could be used, as for instance, a particle landing from above at a random site and stop when the adsorbate is touched. In this case particles would not be rejected, and during one unit of time more than one layer is incorporated to the substrate.

\section{Monte Carlo simulations}

We performed Monte Carlo simulations on square lattices of side $L$, with $L$ ranging from $32$ to $512$, measured in units of the lattice parameter of the substrate. For $L=512$, simulations were done only for particles of size $N=2$ and $N=3$. The simulations were carried out in $(2+1)$ dimensions, and the resulting deposit is a porous three-dimensional structure. We also assumed periodic boundary conditions in both directions of the substrate. As in our simulations we want that one unit of time corresponds to the formation of a new plane of deposited particles parallel to the substrate, it depends on the size of the particle that is being considered for deposition. If we are depositing particles of linear size $N$, the unit of time, which is measured in Monte Carlo steps (MCs), corresponds to $L^{2}/N$ trials of deposition over the surface. According to the growth rules of the model, as established in the last section, depending on the size of the particle to be deposited, during one MCs some particles are not incorporated to the surface, because we do not allow any two particles cross themselves in the same plane. Therefore, a particle that is not added to the surface is reflected, and the trial of deposition is lost. The algorithm for the random deposition of particles is the following: first a cell on a plane parallel to the substrate is randomly selected, and this cell position is defined as the midpoint of the next particle to be deposited. Then, we also choose randomly, the direction, $x$ or $y$, for the deposition. The particle falls horizontally, and is aggregated only if there is enough space around the cell in the selected direction. Otherwise, as established in the model description section, the trial fails and the particle is not incorporated to the substrate.

We start the deposition process at time $t=0$, which corresponds to an initially flat substrate. We record the interface width $w(L,t)$, as a function of time for different values of the side $L$ of the square lattice and different particle sizes $N$. We also considered deposition of particles of different sizes with a statistical weight selected from a Poisson distribution as we have done in our previous work \cite{forgerini_figueiredo}. To get reliable results we averaged over a large number of samples. For the smallest lattice considered in this study $100$ samples are sufficient to get a good statistics.

\section{Numerical results}

We exhibit in Fig. 1 the log-log plot of the interface width as a function of time for the deposition of N-mers, with $N=8$, on a square substrate of side $L=256$. We observe that the time dependence of the roughness can be divided in three different regimes. At the initial times of deposition, less than $30$ MCs, the growth exponent $\beta_{1}$ is close to that of the random deposition of particles where spatial correlation between cells can be neglected. At intermediate times, in the range from $100$ to $800$ MCs, the growth exponent exhibits a completely different behavior from the one observed in one dimension~\cite{forgerini_figueiredo,fabio_2,fabio_silveira}. The correlation between sites becomes important due to the effect of excluded area that a given deposited particle imposes to the others. However, at long times, more than $3000$ MCs in Fig. 1, the number of particles that are reflected from the surface stabilizes, and we reach a constant value for the interface width. This peculiar behavior was found in the electrophoretic deposition of polymer chains on a two-dimensional substrate~\cite{bentrem}.

\begin{figure}
 \centering
 \includegraphics[width=0.7\textwidth]{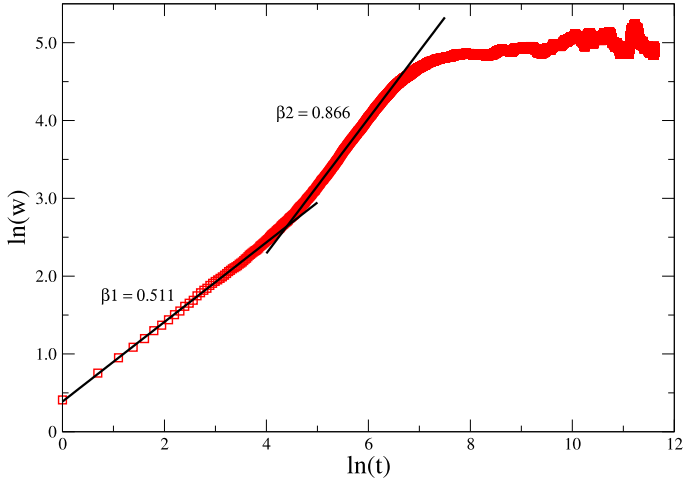}
 \caption{(Color online) Log-log plot of the interface width versus time. Deposition of particles of size $N=8$ on a square lattice of side $L=256$. The values of the growth exponents $\beta_{1}$ and $\beta_{2}$ are indicated in the figure.}
\end{figure}

We show in Fig. 2 the behavior of the interface width for the deposition on a square substrate of side $L=128$ and different particle sizes. In Fig. 2d we are plotting the interface width when we consider the deposition of a mixture of particles, with sizes in the range from $1$ to $15$, which are randomly selected from a Poisson distribution like one used in our previous work. Except for $N=1$, the interface width saturates as shown in Fig. 2, and we observe the same trends as seen in Fig. 1: at the initial times we have a growth exponent $\beta_{1}$, very close to $1/2$, typical of uncorrelated growths, a new exponent $\beta_{2}$ at intermediate times, which depends on the size of the particle being deposited, and finally, the interface width saturates at long times of deposition.

\begin{figure}
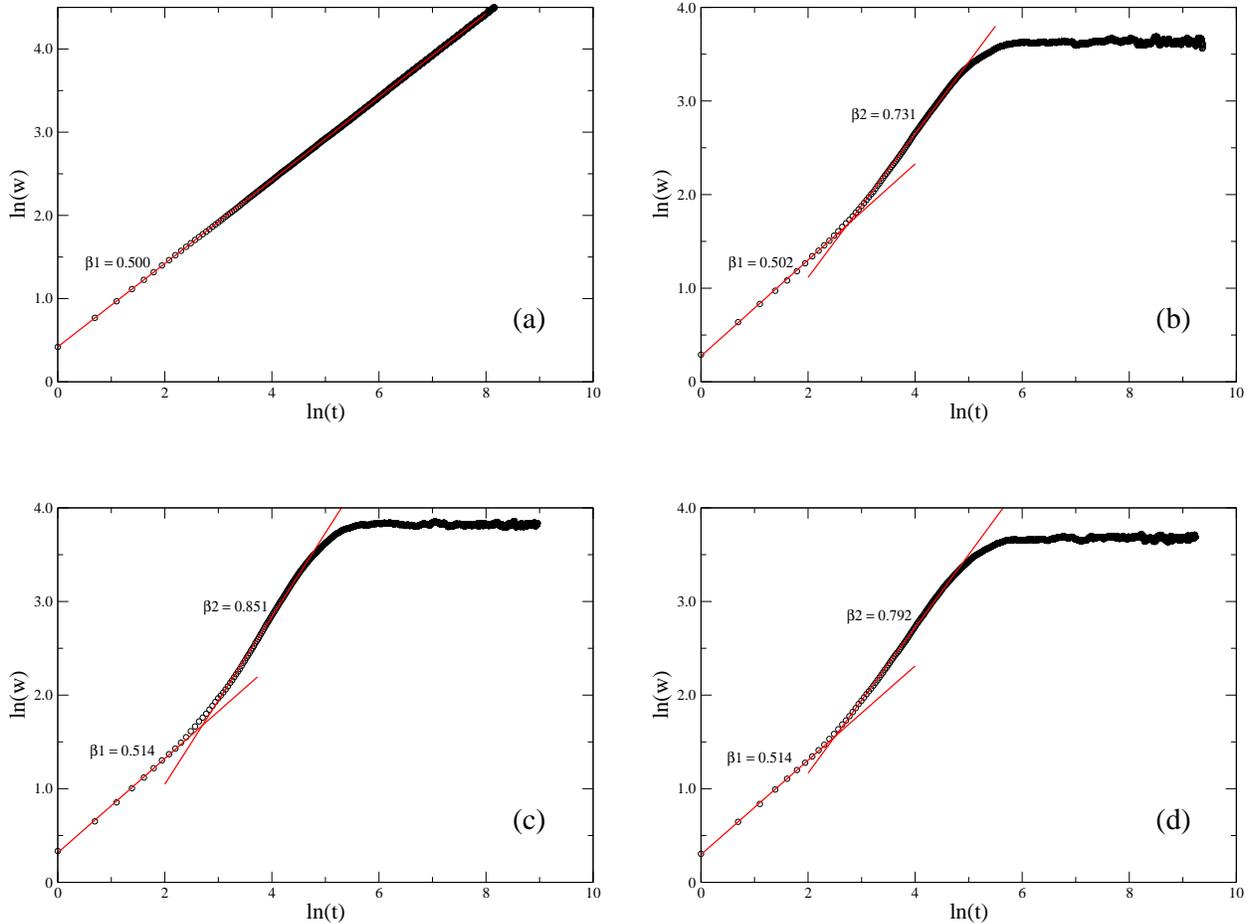

    \vskip 0.2cm
    \begin{minipage}[t]{0.48 \linewidth}
	\includegraphics[width=\linewidth]{new_fig2a.eps}\\
    \end{minipage}\hfill
    \begin{minipage}[t]{0.48 \linewidth}
	\includegraphics[width=\linewidth]{new_fig2b.eps}\\
    \end{minipage}\hfill
    \vskip 1cm
    \begin{minipage}[t]{0.48 \linewidth}
	\includegraphics[width=\linewidth]{new_fig2c.eps}\\
    \end{minipage}\hfill
    \begin{minipage}[t]{0.48 \linewidth}
	\includegraphics[width=\linewidth]{new_fig2d.eps}\\
    \end{minipage}\hfill
 \caption{(Color online) Log-log plots of the interface width for the deposition of N-mers on a square lattice of side $L=128$. The values of the growth exponents $\beta_{1}$ and $\beta_{2}$ are indicated in the plots. (a) $N=1$, (b) $N=6$, (c) $N=9$ and (d) mixture of particles with sizes $1\leq N\leq15$.}
 \label{fig:roughness}
\end{figure}

Now we turn our attention to the dependence of the growth exponent $\beta_{2}$ and the roughness exponent $\alpha$ on the length of particles. For each particle size we observe a saturation of the interface width, which depends on the side $L$ of the square lattice. At very long times, we expect that roughness scales with the substrate size as $w_{sat}\sim L^{\alpha}$. The exponent $\alpha$ can be estimated from $w_{sat}$ after we extrapolate the effective exponents defined by the equation \cite{jonatas_fabio}

\begin{equation}
\alpha (L) \equiv\frac{\ln[w_{sat}(L)/w_{sat}(L/2)]}{\ln2}.
\end{equation}

In Fig. 3 we plot the exponent $\alpha(L)$ as a function of $L^{-1}$ for three different particle sizes. When we fit the data points, we find, for large values of $L$, a value close to $1/3$. Within the error bars, we assume that this is the value of the exponent $\alpha$ for all the particle sizes considered in this study, except for $N=1$, where it is not defined. On the other hand, we plot in Fig. 4, the exponent $\beta_{2}$, which describes
the time evolution of the interface width at intermediate times, as a function of the particle size, and for three different linear sizes of the substrate. The dependence of $\beta_{2}$ on $L$ is very weak, while it depends strongly on the size of the particle. This is evident when the size of particle changes from $N=2$ to $N=3$. For $N=2$ the growth exponent $\beta_{2}$ is smaller than $\beta_{1}$, while it is larger than $\beta_{1}$ for $N\geq3$. As to be expected, the coverage by dimers is more effective than the corresponding coverage by linear N-mers larger than three units of the underlying lattice parameter. For instance, while the jamming coverage for dimers~\cite{dickman} is $0.90654$, the corresponding value for trimers~\cite{bonnier} is $0.8470$.

\begin{figure}
 \centering
 \includegraphics[width=0.7\textwidth]{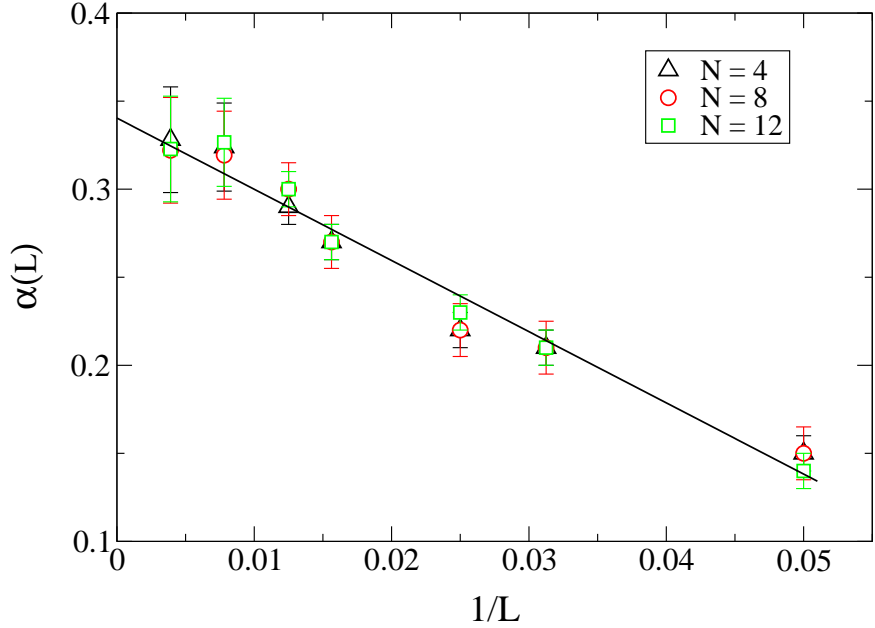}
 \caption{(Color online) Roughness exponent $\alpha$ as a function of $L^{-1}$ for three different particle sizes as indicated in the figure. From the fit we find that $\alpha\approx 1/3$ for large values of $L$.}
\end{figure}

\begin{figure}
 \centering
 \includegraphics[width=0.7\textwidth]{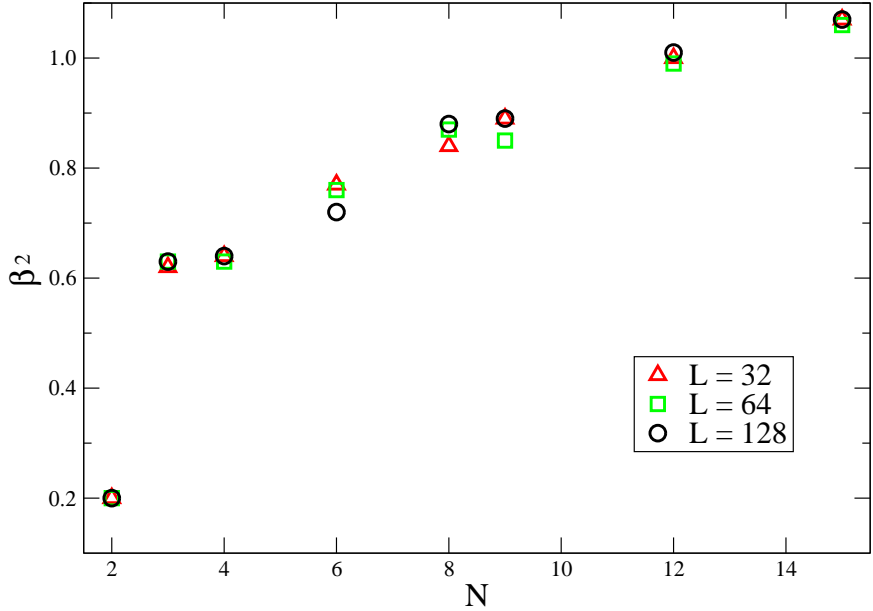}
 \caption{(Color online) The growth exponent $\beta_{2}$ as a function of the particle length $N$ for three different sides $L$ of the square lattice as indicated in the figure.}
\end{figure}

We display in Table I the data for the deposition of particles with sizes in the range $2$ to $15$, as well as for the deposition of a mixture of particles, onto a square substrate of side $L=128$. At the initial growth stage denoted by the growth exponent $\beta_{1}$, the deposition is clearly driven by an uncorrelated mechanism. However, at intermediate times, we observe a sudden jump in the value of the exponent $\beta_{2}$ when the particle size changes from $N=2$ to $N=3$. This behavior appears to be due to the increase in the spatial correlations with the size of the particles. The excluded area formed around each deposited particle increases with its size. The increase in the spatial correlations is more striking when we consider the sizes $N=1$ and $N=2$. When we go from $N=1$, for which we always have $\beta_2 = \beta_1 $ close to $1/2$ at any time, with no spatial correlations, to $N=2$, the exponent $\beta_2$, measured at intermediate times, decreases to a value close to $0.20$.

We also collected the figures of the saturation value of the interface width for the same set of particle sizes. As it is almost constant, it corroborates the data of Fig. 3, where the same fact was observed for all lattice sizes considered in this work. Therefore, by assuming that $\alpha = 1/3$, and taking the values of $\beta_{2}$ from Table I, we can estimate the values of the dynamical critical exponent $z$ through the relation $z=\alpha/\beta_{2}$, which are also displayed in Table I. As to be expected, the values we find for the deposition of a mixture of particles are intermediary between those determined for the smallest and largest particles. Although we have considered deposition of particles as large as $N=15$, Fig. 4 shows that the rate of increasing the value of the exponent $\beta_{2}$ is lower for higher values of $N$. The same trend was observed in simulations performed in one dimension~\cite{forgerini_figueiredo}. Contrary to one-dimensional results, where the scaling exponents were related to the Villain-Lai-Das Sarma equation, in the case of deposition of particles of different sizes in two dimensions, the growth exponents are non-universal. As far as we know, there is no continuum growth equation that can describe this non-trivial model for deposition of N-mers in two dimensions.

\begin{table}[htb!]
 \centering \vskip 0.2cm \begin{normalsize} \begin{tabular}{ccccc}
\hline 
Chain Size (N)  & $\beta_{1}$  & $\beta_{2}$  & ln($w_{sat}$)  & $z$ \tabularnewline
\hline 
%\hline
2  & 0.50$\pm$0.05  & 0.20$\pm$0.05  & 3.6$\pm$0.9  & 1.67$\pm$0.42 \tabularnewline
%\hline
3  & 0.50$\pm$0.03  & 0.63$\pm$0.04  & 3.7$\pm$0.4  & 0.53$\pm$0.03 \tabularnewline
%\hline
4  & 0.49$\pm$0.07  & 0.63$\pm$0.04  & 3.8$\pm$0.9  & 0.53$\pm$0.03 \tabularnewline
%\hline
6  & 0.50$\pm$0.02  & 0.73$\pm$0.02  & 3.7$\pm$0.6  & 0.45$\pm$0.01 \tabularnewline
%\hline
8  & 0.50$\pm$0.03  & 0.83$\pm$0.03  & 3.9$\pm$0.9  & 0.37$\pm$0.01 \tabularnewline
%\hline
9  & 0.51$\pm$0.04  & 0.85$\pm$0.05  & 3.9$\pm$0.9  & 0.39$\pm$0.02 \tabularnewline
%\hline
12  & 0.50$\pm$0.06  & 0.99$\pm$0.04  & 4.0$\pm$0.6  & 0.34$\pm$0.01 \tabularnewline
%\hline
15  & 0.50$\pm$0.08  & 1.06$\pm$0.06  & 4.0$\pm$0.8  & 0.32$\pm$0.02 \tabularnewline
%\hline
Mixture  & 0.51$\pm$0.05  & 0.79$\pm$0.09  & 3.9$\pm$0.2  & 0.42$\pm$0.03 \tabularnewline
\hline
\end{tabular}
\end{normalsize} 
\caption{Growth exponents $\beta_{1}$ and $\beta_{2}$, saturation value of the interface width, and the dynamic exponent $z$ for the deposition of $N-mers$ and for a mixture (last line) of $N-mers$ with sizes in the range from $N=1$ to $N=15$ onto a square lattice of side $L=128$.}
\end{table}

\section{Conclusions}

We have considered in this work a growth model for the deposition of linear N-mers, with $1\leq N \leq15$, over a square lattice. The particles are one unit height, and are randomly deposited in two perpendicular directions parallel to the substrate. By employing Monte Carlo simulations, we calculated the interface width as a function of time. We have shown that three different regimes emerge from calculations. At the initial times, the behavior is typical of an uncorrelated growth, given by the exponent $\beta_{1}\approx1/2$,
for whatever particle size $N$. At intermediate times, the growth exponent is no more universal. While for $N=2$ its value is $\beta_{2} \approx0.20$, it jumps to $\beta_{2}\approx0.60$ for $N=3$, showing that spatial correlations increase with particle size. This happens because the excluded region around each deposited particle increases with its linear size $N$. For $N>3$ the rate of increasing the value of $\beta_2$ is lower for higher values of $N$. Finally, at very long times and for very large lattices, the interface width attains a constant value $\alpha \approx 1/3$, which is independent of the length $N$ of the deposited particle. Despite the same deposition model in one dimension is described by the Villain-Lai-Das Sarma continuum equation, in two dimensions it is clearly non-universal, with growth and dynamical exponents depending on the length of
the deposited particles.

\begin{acknowledgements}
The authors would like to thank Conselho Nacional de Desenvolvimento Cient\'{i}fico e Tecnol\'ogico (CNPq) for the financial support.
\end{acknowledgements}


\begin{thebibliography}{28}

\bibitem{barabasi} A. Barab\'asi and H. E. Stanley, {\textit{Fractal
Concepts in Surface Growth}}. Cambridge University Press, Cambridge,
1995.

\bibitem{meakin} P. Meakin, {\textit{Fractals, Scaling and Growth
far from Equilibrium}}. Cambridge University Press, Cambridge, 1998.

\bibitem{FV} F. Family and T. Vicsek, J. Phys. A \textbf{18}, L75
(1985).

\bibitem{EW82} S. F. Edwards and D. R. Wilkinson, Proc. R. Soc. London
A \textbf{381}, 17 (1982).

\bibitem{KPZ86} M. Kardar, G. Parisi and Y. Zhang, Phys. Rev. Lett.
\textbf{56}, 889 (1986).

\bibitem{Vold} M. J. Vold, J. Colloid Sci. \textbf{14}, 168 (1959).

\bibitem{reis_stafiej} F. D. A. Aar\~ao Reis and Janusz Stafiej, Phys.
Rev. E \textbf{76}, 011512 (2007).

\bibitem{Fernando} B. A. Mello, A. S. Chaves, and F. A. Oliveira,
Phys. Rev. E \textbf{63}, 041113 (2001).

\bibitem{das_sarma} S. Das Sarma and P. Tamborenea, Phys. Rev. Lett.
\textbf{66}, 325 (1991).

\bibitem{albano} E. V. Albano, R. C. Salvarezza, L. Vazquez, and
A. J. Arvia, Phys. Rev. B \textbf{59}, 7354 (1999).

\bibitem{drossel} B. Drossel and M. Kardar, Phys. Rev. Lett. \textbf{85},
614 (2000).

\bibitem{horowitz_arvia} C. M. Horowitz, M. A. Pasquale, E. V. Albano,
and A. J. Arvia, Phys. Rev. B \textbf{70}, 033406 (2004).

\bibitem{Reis} F. D. A. Aar\~ao Reis, Phys. Rev. E \textbf{73}, 021605
(2006).

\bibitem{barato_oliveira} A. C. Barato and M. J. Oliveira, J. Phys.
A \textbf{40}, 8205 (2007).

\bibitem{cerdeira_95} W. Wang and H. A. Cerdeira, Phys. Rev. E \textbf{52},
6308 (1995).

\bibitem{elnassar_cerdeira} H. F. El-Nashar and H. A. Cerdeira, Phys.
Rev. E \textbf{61}, 6149 (2000).

\bibitem{karmakar} R. Karmakar, T. Dutta, N. Lebovka, and S. Tarafdar,
Physica A \textbf{348}, 236 (2005).

\bibitem{trojan} K. Trojan and M. Ausloos, Physica A \textbf{326},
492 (2003).

\bibitem{caglioti} E. Caglioti, V. Loreto, H. J. Herrmann, and M.
Nicodemi, Phys. Rev. Lett. \textbf{79}, 1575 (1997).

\bibitem{bentrem} Frank W. Bentrem, R. B. Pandey, and Fereydoon Family,
Physical Review E, 62, 914 (2000).

\bibitem{seung-woo} S.-W Son, M. Ha and H. Jeong, J. Stat. Mech.
\textbf{P02031} (2009).

\bibitem{gangshi} Gangshi Hu, Jianqiao Huang, Gerassimos Orkoulas,
and Panagiotis D. Christofides, Phys Rev. E \textbf{80}, 041122 (2009).

\bibitem{forgerini_figueiredo} F. L. Forgerini and W. Figueiredo,
Phys. Rev. E \textbf{79}, 041602 (2009).

\bibitem{fabio_2} F. D. A. Aar\~ao Reis, Physica A \textbf{364}, 190
(2006).

\bibitem{fabio_silveira} F. A. Silveira and F. D. A. Aar\~ao Reis,
Phys. Rev. E \textbf{75}, 061608 (2007).

\bibitem{jonatas_fabio} J\^onatas A. R. Euz\'ebio and F. D. A. Aar\~ao
Reis, Phys. Rev. E \textbf{80}, 021605 (2009). 

\bibitem{dickman} M. J. de Oliveira, T. Tom\'e and R. Dickman, Phys.
Rev. A 46, 6294 (1992).

\bibitem{bonnier} B. Bonnier, M. Hontebeyrie, Y. Leroyer, C. Meyers
and E. Pommiers, Phys. Rev. E 49, 305 (1994).

\end{thebibliography}
\end{document}